\documentclass[12pt]{article}

\usepackage{graphicx}
\usepackage{amsmath}
\usepackage{amssymb}

\usepackage{color}
\input{colordvi.tex}

\def\be{\begin{eqnarray}}
\def\ee{\end{eqnarray}}
\def\ne{\nonumber\end{eqnarray}}

\def\d{\partial}
\def\pd#1{{\partial\over\partial{#1}}}
\def\abs#1{\left|{#1}\right|}
\def\hf{{1 \over 2}}
\def\rtni{\sqrt{2}}
\def\Mp{M_{\rm p}}
\def\bN{\bar{N}}
\def\z{\zeta}
\def\CO{{\cal O}}
\newcommand{\lmk}{\left(}  \newcommand{\rmk}{\right)}

\newcommand{\gsim}{\mathrel{\mathop{\kern-0pt >}\limits_{\sim}}}
\newcommand{\lsim}{\mathrel{\mathop{\kern-3pt <}\limits_{\sim}}}
\def\sitarel#1#2{\mathrel{\mathop{\kern-0pt #1}\limits_{#2}}}

\begin{document}

\vspace*{-1cm}\hspace*{12cm}
\hfill\parbox{4cm}
{\normalsize 
{UT-08-21}\\
{July, 2008}
}\\

\vskip .4in
\centerline{\Large\bf 
                 Reheating in Chaotic D-Term Inflation
}

\vskip .4in
\centerline{\sc Teruhiko Kawano$^{\star}$
                and 
                Masahide Yamaguchi$^{*,\dagger}$}

\vskip .2in
\centerline{\it ${}^{\star}$Department of Physics, University of Tokyo, Hongo, 
Tokyo 113-0033, Japan}
\vskip 0in
\centerline{\it ${}^{*}$Department of Physics and Mathematics, 
Aoyama Gakuin University}
\centerline{\it Sagamihara 229-8558, Japan}
\vskip 0in
\centerline{\it ${}^{\dagger}$Department of Physics, Stanford
University, Stanford CA 94305}


\vskip .4in 
\centerline{\small \bf Abstract} 
\medskip 
\noindent{\small 
A simple model is discussed to give rise to successful reheating in
chaotic D-term inflation with a quadratic inflaton potential,
introducing a trilinear coupling in the K\"ahler potential.
Leptogenesis through the inflaton decay is also discussed in this
model.}

\vskip .5in
\section{Introduction}

Inflation gives a natural solution to the horizon and the flatness
problems as well as the mechanism to generate primordial density
fluctuations \cite{Inflation}. The recent observations like the cosmic
microwave background anisotropy by the Wilkinson Microwave Anisotropy
Probe (WMAP) \cite{Komatsu:2008hk} strongly supports inflationary
scenarios.  Among many types of inflation models proposed thus far,
chaotic inflation, which was proposed by Linde \cite{Linde:1983gd}, is
attractive in that it does not suffer from any initial condition
problem. Since supersymmetry is a plausible solution to the hierarchy
problem in particle physics, it is natural to consider inflation models
in supergravity.  However, it is very difficult to realize chaotic
inflation in supergravity.  This is mainly because the F-term potential
in supergravity has the exponential of the K\"ahler potential as a
factor, and in general prevents scalar fields from having an initial
value much larger than the reduced Planck scale $\Mp \simeq 2.4 \times
10^{18}$~GeV.

One method to circumvent this difficulty is to introduce the shift
symmetry
\cite{Kawasaki:2000yn,Kawasaki:2000ws,Yamaguchi:2000vm,Yamaguchi:2001pw,
Yamaguchi:2001zh}, which guarantees the flat directions especially in
the K\"ahler potential and allows a field along the flat directions to
be identified with an inflaton field.\footnote{See
Refs. \cite{Goncharov:1983mw,Goncharov:1985ka,Murayama:1993xu} for early
attempts to realize chaotic inflation in supergravity. See also
Ref. \cite{Silverstein:2008sg}, in which the authors attempt to realize
chaotic inflation in superstring by use of the monodromy.} Another
method is to use the D-term potential, which does not have the
exponential factor as the F-term potential does \cite{KY}. A field along
the F-flat direction, which is automatically lifted by the D-term
potential, can play the role of inflaton to give rise to chaotic
inflation.  The idea was used in the papers \cite{KY,KKY} to give
chaotic inflation models with a quartic potential. Chaotic inflation
models with a quartic potential in general seems disfavored by the
recent observations \cite{Komatsu:2008hk} and may require additional
contributions like curvaton
\cite{Moroi:2005kz,Moroi:2005np,Ichikawa:2008iq} to explain the
observations.

Recently, a simple chaotic inflation model with a quadratic potential
was proposed in supergravity \cite{FI}, which is based on ${\cal N}=1$
supersymmetric $U(1)$ gauge theory where the Fayet-Iliopoulos parameter
replaced by a dynamical field plays the role of inflaton. Since the
potential of the model has only the D-term potential and depends on the
Fayet-Iliopoulos parameter quadratically, the model gives rise to
chaotic inflation with a quadratic potential of the inflaton
field. Thus, in the paper \cite{FI}, the eta-problem was naturally
solved. It is well-known that many models in supergravity for inflation
suffer from not only the eta-problem but also the overproduction of
gravitino.  However, to avoid the overproduction of gravitino, one needs
to make sure two necessary conditions; the first is to obtain
sufficiently low reheating temperature, and the second is to prevent the
inflaton field from developing the vacuum expectation value (VEV) after
inflation.  It is also known that it is extremely difficult to meet
these two conditions. Therefore, in this paper, we concentrate on the
reheating stage in the model. By introducing interactions of the
inflaton with right-handed neutrinos, it will be discussed that one can
achieve successful reheating with low reheating temperature and
leptogenesis.

In the next section, after briefly reviewing the model of
Ref. \cite{FI}, we introduce an interaction necessary for reheating and
investigate its effect on the dynamics of the inflation. In
Sec. \ref{sec:reheating}, we evaluate the reheating temperature and show
that leptogenesis really works in our scenario. In the final section, we
give summary and discussions.

\section{Model and Chaotic Inflation}
\label{sec:model}

The model proposed in the paper \cite{FI} is ${\cal N}=1$ supersymmetric
$U(1)$ gauge theory with a chiral superfield $S$, whose scalar component
is denoted by
\be
S={1\over\sqrt{2}}\left(\rho+i\zeta\right).
\ne
The real scalar field $\rho$ plays the role of the Stuckelburg field
giving mass to the $U(1)$ gauge multiplet, while the real scalar field
$\zeta$ is the inflaton field appearing in the D-term potential as if it
was the Fayet-Iliopoulos parameter.

In order to achieve reheating in the model, we also introduce neutral
chiral superfields $N_i$ ($i=1,2,3$), which play the role of
right-handed (s)neutrinos, and the minimal supersymmetric standard model
(MSSM) superfields $\phi^k$. Under the gauge transformation, they
transform as
\be
V \to V+i\left(\Lambda-\bar\Lambda\right), \qquad 
S \to S+\rtni M\Lambda, \qquad
N_i \to N_i, \qquad
\phi^k \to \phi^k, \qquad
\ne
where $V$ is the vector superfield of the gauge field $v_\mu$, and $M$
will turn out to be the inflaton mass.\footnote{The notation in this
paper is slightly different from the one in \cite{FI}, and in this paper
we follow the convention of chapter XXIV and Appendix G in the textbook
\cite{WB}, with trivial exceptions for the K\"ahler potential and the
superpotential. } Since we take the K\"ahler potential
\be
K=
-\hf\left(S-\bar{S}+\rtni\,iM\, V\right)^2
-i{g\over\Mp}\left(S-\bar{S}+\rtni\,iM\, V\right)\bN_iN_i
+\bN_iN_i
+\hat{K}(\phi_k^{*},\phi^k),
\label{Kahler}
\ee
with $g$ a coupling constant, it follows that the Killing potential $D$
is given by
\be
D={M\over\rtni}\left[-i\left(S-\bar{S}\right)+{g\over\Mp}\bN_iN_i\right]. 
\ne
Note that such a trilinear coupling between the inflaton and the
right-handed neutrinos in the K\"ahler potential is peculiar to our
model because the R-symmetry prohibits such a term for many existing
models.  Furthermore, we take the superpotential to be
\be
W={m_{i}\over2}N_iN_i+\hat{W}(N,\phi^k),
\ne
where $\hat{W}(N,\phi^k)$ includes the Yukawa interactions $y_{ij} N_i
L_j H_u$ of the right handed neutrinos $N_i$ with the left-handed lepton
superfields $L_i$ ($i=1,2,3$) and the Higgs superfield $H_u$.

Since it is convenient to describe the model in terms of gauge invariant
variables, let us change the variable as
\be
v_{\mu} \to v_{\mu}+{1\over{M}}\d_{\mu}\rho 
\ne
to completely eliminate the field $\rho$ from the action as the gauge
degrees of freedom.  One then finds that the potential is given by
\be
V &=&
\exp\left[{1\over\Mp^2}\left(\z^2+\abs{N_i}^2+{\rtni{g}\over\Mp}\z\abs{N_i}^2
                             +\hat{K}(\phi_k^{*},\phi^k)
                       \right)\right] \times  \nonumber \\
  && \qquad \biggl[ 
       \frac{1}{1+\frac{\sqrt{2}g}{\Mp}\zeta} |D_{N_i} W|^2
      +\frac{1+\frac{\sqrt{2}g}{\Mp}\zeta}{1+\frac{\sqrt{2}g}{\Mp}\zeta
             -\frac{g^2}{\Mp^2}|N_{i}|^2}
       \left| \frac{\sqrt{2} \zeta}{\Mp^2}W - \frac{g}{\Mp}
              \frac{1}{1+\frac{\sqrt{2}g}{\Mp}\zeta}
              (m_{i}N_{i}^2+\hat{W})  
       \right|^2 \biggr. \nonumber \\
  && \qquad \,\,   \biggl.
      +g^{k\bar{l}}D_k\hat{W}(D_l\hat{W})^{*}
      -\frac{3}{\Mp^2}|W|^2
    \biggr] \nonumber \\
  && + \frac{M^2}{2} \left( \zeta+\frac{g}{\sqrt{2}\Mp}|N_{i}|^2 \right)^2
     + V_D^{\rm MSSM}. 
\label{V}
\ee
where $D_kW$ denotes 
\be
D_kW=\pd{\phi^k}W+{1\over\Mp^2}\left(\pd{\phi^k}K\right)W.
\ee
Note that the first term on the right hand side of Eq.~(\ref{V}) is the
F-term potential, and the remaining terms the D-term potential.
$V_D^{\rm MSSM}$ represents the contribution coming from the standard
model gauge group to the D-term potential.

When the universe starts around the Planck energy density and the
inflaton $\zeta$ takes a value much larger than $\Mp$, 
the exponential factor in the F-term potential enforces the conditions
$D_{N_i} W = D_kW = W = 0$ to set the contribution from the F-term in 
the potential to zero.
Since the directions with $N_i=0$ satisfy the condition, 
the system could remain in one of the directions during inflation.
In fact, the effective masses squared of the neutrinos at
the origin are estimated as
\be
m_{i}^2(\z) \simeq \frac{m_{i}^2}{1+\frac{\sqrt{2}g}{\Mp}\zeta}
               e^{{\z^2\over\Mp^2}}
               +{g\z\over\rtni\Mp}M^2.
\ne
On the other hand, the Hubble parameter squared in the same period is
given by
\be
H^2 ~\simeq~ {1\over6}\left({M\over\Mp}\right)^2\z^2. 
\ne
Therefore, for large $\zeta$, the effective masses of the neutrinos are
much larger than the Hubble parameter so that $N_{i}$ remains at the
origin during inflation\footnote{ Since the kinetic terms of the
neutrinos $N_i$ are not canonical, we must have taken account of it to
estimate their effective masses. However, the exponential factor is so
large that such an effect is negligible.}. Here and hereafter, for
simplicity, we will take positive $g$ without loss of generality and
assume positive $\z$ during inflation. Thus, the potential of the
inflaton is dominated by the D-term with the quadratic potential, which
leads to chaotic inflation. One thus needs to take the inflaton mass $M$
to be of order $10^{13}$~GeV in order to explain the primordial density
fluctuations.

\section{Reheating and Leptogenesis}
\label{sec:reheating}

So thus, we have seen that the model of the paper \cite{FI} with the
neutrino fields $N_i$ still gives rise to successful chaotic inflation.
In this section, we would like to discuss reheating and the baryon
asymmetry of the universe through the leptogenesis scenario
\cite{Fukugita:1986hr} via the inflaton decay
\cite{Kumekawa:1994gx,Lazarides:1999dm,Giudice:1999fb,Asaka:1999yd,Asaka:1999jb}
in our model.

After inflation, the inflaton field starts oscillating about the
origin.\footnote{The kinetic term of the neutrinos become singular at
$\zeta = -\Mp/(\sqrt{2} g)$. However, since we assume that the inflaton
takes a positive value during inflation, it cannot reach this value for
$g \ll 1$ during the oscillation phase as well.}  Thanks to the
interaction term between the inflaton $\zeta$ and the neutrino
multiplets $N_i$ in the K\"ahler potential, the inflaton can decay into
the (s)neutrinos through the interactions
\be
{g\over\rtni\Mp}\zeta\left[-M^2\abs{N_i}^2
+(2m_i^2)\abs{N_i}^2
+\bar{N}_i\d_{\mu}\d^{\mu}{N}_i+{N}_i\d_{\mu}\d^{\mu}\bar{N}_i
-i\bar\chi_i\bar\sigma^{\mu}\d_{\mu}\chi_i
-i\chi_i\sigma^{\mu}\d_{\mu}\bar\chi_i\right],
\label{decay}
\ee
where the field $\chi_i$ is the spinor component of the chiral
superfield $N_i$. Here you should notice that the trilinear coupling of
the first term originates from the D-term potential, which is peculiar
to our D-term model of chaotic inflation. In the last four terms on the
right hand side of Eq. (\ref{decay}), the derivatives can be replaced by
the neutrino mass $m_i$ by using the on-shell condition in lowest order
perturbation theory.

Assuming that $m_1$ is of order $10^{12}$~GeV, while $m_2$ and $m_3$ are
assumed to be larger than the inflaton mass $M$, we only have to
consider the decay of the inflaton into $N_{1}$.  Therefore, the first
term coming from the D-term potential gives the dominant contribution to
the decay rate of the inflaton.  Then, one obtains the decay rate of the
inflaton $\z$
\be
\Gamma_{\zeta}~\simeq~{1\over16\pi{M}}\left({gM^2\over\rtni\Mp}\right)^2. 
\ne
Therefore, the reheating temperature is given by
\be
T_{R} ~\simeq~ 3.0 \times 10^7~{\rm GeV} 
               \lmk \frac{g}{0.1} \rmk
               \lmk \frac{M}{10^{13}~\mbox{GeV}}\rmk^{3\over2}.
\ne
Inflation models with such a low reheating temperature are viable for
the wide range of the gravitino mass \cite{Kawasaki:2008qe}.

The produced (s)neutrinos $N_{i}$ decay into (s)leptons $L_{j}$ and
Higgs(ino) doublets $H_{u}$ through the Yukawa interactions 
\be
  W = h_{\nu}^{ij}N_{i}L_{j}H_{u} 
\ee
in the superpotential, where we have taken a basis where the mass matrix
for $N_{i}$ is diagonal. We also assume $|(h_{\nu})_{i3}| >
|(h_{\nu})_{i2}| \gg |(h_{\nu})_{i1}|$ (i = 1, 2, 3). We consider only
the decay of $N_{1}$ with the above assumption that the mass $m_1$ is
much smaller than the others.  The interference between the tree-level
and the one-loop diagrams including vertex and self-energy corrections
generates the lepton asymmetry
\cite{Fukugita:1986hr,Covi:1996wh,Flanz:1994yx,Buchmuller:1997yu},
\be
   \label{eqn:cpasym}
  \epsilon_1 
    &\equiv&
    \frac{ \Gamma (N_1 \rightarrow H_u + l )
         - \Gamma (N_1 \rightarrow \overline{H_u} + \overline{l} ) }
         { 
\Gamma (N_1 \rightarrow H_u + l )
         +\Gamma (N_1 \rightarrow \overline{H_u} + \overline{l} ) 
}
    \nonumber \\
    &\simeq& - 
    \frac{3}{ 8 \pi \left( h_\nu h_\nu^{\dagger} \right)_{11} }
    \sum_{i=2,3} 
        \mbox{Im} \left( h_\nu h_\nu^{\dagger} \right)_{1i}^2 
        \frac{{m_1}}{{m_i}}
    \simeq \frac{3}{8\pi} \frac{{m_1}}{\langle H_u\rangle^2}m_{\nu_3}
\delta_{\rm eff}
\nonumber 
\\
&\sim&10^{-4} \left(\frac{{m_1}}{10^{12}~\rm GeV}\right)
\left(\frac{m_{\nu_3}}{10^{-2}~\rm eV}\right) \delta_{\rm eff},
\ee
with the effective CP violation phase given by
\be
  \delta_{\rm eff}\equiv -\frac{ \mbox{Im} 
     \left[ h_\nu (m^*_{\nu}) h_\nu^{T} \right]_{11}^2 }
    {m_{\nu_3}\left( h_\nu h_\nu^{\dagger} \right)_{11}},
\ee
where $m_{\nu_{3}}$ is a mass eigenvalue of the left-handed 
neutrino mass matrix $m_{\nu}$ and can be estimated by the seesaw mechanism
\cite{seesaw,GPS} as
\be
  m_{\nu_{3}} &\simeq& \frac{\left|\left( h_\nu \right)_{33}\right|^2
                           \langle H_{u} \rangle^{2}}{m_{3}} \nonumber \\
              &\sim& 10^{-2}~{\rm eV} \left(
                       \frac{\left|\left( h_\nu\right)_{33}\right|}
                            {10^{-1}} \right)^2
                        \left(\frac{m_{3}}{10^{13}~\rm GeV} \right)^{-1}.
\ee
Here we assumed that the $(h_{\nu})_{33}$ is dominant in $h_{\nu}$ and
$m_3\gg m_1$.

The total decay rate $\Gamma_{N_1}$ of $N_{1}$ is given by
\be
  \Gamma_{N_1} &=& \Gamma (N_1 \rightarrow H_u + l )
         + \Gamma (N_1 \rightarrow \overline{H_u} + \overline{l} )
                   \nonumber \\
               &\simeq& \frac{1}{8\pi} |(h_{\nu})_{13}|^{2} m_{1} 
                         \nonumber \\
               &\sim& 10^6~{\rm GeV} 
                      \left( \frac{|(h_{\nu})_{13}|}
                                {10^{-2}} \right)^{2}
                      \left( \frac{m_{1}}{10^{12}~\rm GeV} \right).
\ee
Thus, the decay rate $\Gamma_{N_1}$ is much larger than the decay rate
of the inflaton $\Gamma_{\zeta}$ so that the produced $N_{1}$
immediately decays into lepton and Higgs supermultiplets once they are
produced from the inflaton.

A part of the produced lepton asymmetry is converted into the baryon
asymmetry through the sphaleron processes, which can be estimated
\cite{Khlebnikov:1988sr,Harvey:1990qw} as
\be
  \frac{n_B}{s} \simeq - \frac{8}{23} \frac{n_L}{s},
\ee
where we have assumed the standard model with two Higgs doublets and
three generations. In order to explain the observed baryon number
density
\be
  \frac{n_B}{s} \simeq (0.4 - 1) \times 10^{-10},
\ee
it is necessary to achieve the lepton asymmetry
\be
  \frac{n_L}{s} \simeq - (1 - 3) \times 10^{-10}.
\ee

Now we estimate the lepton asymmetry produced through the inflaton
decay. For $m_{1} \gtrsim 10^{12}$~GeV, $m_{1}$ is much larger than the
reheating temperature $T_{R}$. In this case, the produced $N_{1}$ is out
of equilibrium and the ratio of the lepton number to entropy density can
be estimated as
\be
  \frac{n_L}{s} 
          &\simeq&
       \frac32~\epsilon_1 B_r \frac{T_R}{{M}} 
    \nonumber \\
    &\sim& 
    - 5 \times 10^{-10} \delta_{\rm eff} B_r
    \lmk \frac{g}{10^{-1}} \rmk
    \lmk \frac{M}{10^{13}~\mbox{GeV}} \rmk^{{\frac12}}
    \lmk \frac{m_1}{10^{12}~\mbox{GeV}} \rmk,
\ee
where $B_r$ is the branching ratio of the inflaton $\zeta$ into $N_1$.
As stated before, we typically take $m_1$ to be of order $10^{12}$~GeV,
and $m_2$ and $m_3$ to be at least larger than $10^{13}$~GeV, which
implies $B_r =\CO(1)$. Then, the typical values of the parameters can
explain the baryon number density in the present universe. Thus,
sufficient baryon asymmetry can be generated for low reheating
temperature without fine-tuning of the parameters in our model.

\section{Summary and Discussions}

In this paper, we have discussed how to reheat the universe in the model
of the paper \cite{FI}.  For this purpose, we introduced the interactions
of the inflaton with right-handed neutrinos in the K\"ahler potential,
which yields the effective trilinear couplings between them originating
from the D-term potential. This property is peculiar to the model of
D-term chaotic inflation. We have seen that this model leads to
successful chaotic inflation even after introducing such an
interaction. The inflaton can decay into right-handed (s)neutrinos via
the effective trilinear interactions.\footnote{The inflaton could also
couple to other particles if similar trilinear couplings were introduced
in the K\"ahler potential.  However, the branching ratio to the right
handed neutrinos would be dominant in the inflaton decay, as long as the
coupling constant to them is not significantly suppressed in comparison
to those to the other particles.  Therefore, they would hardly affect
our results on the reheating temperature.} The produced right-handed
(s)neutrinos quickly decay into the MSSM particles through the Yukawa
interactions so that the universe is completely reheated. We have found
the typical reheating temperature to be $10^{7}$~GeV for $g \simeq 0.1$,
which is low enough to avoid the overproduction of gravitino for a wide
range of the gravitino mass. The decay of right-handed (s)neutrinos
produces the lepton asymmetry if the CP violation exists. We have shown
that the amount of the produced lepton asymmetry is sufficient to
explain the present baryon asymmetry for typical parameter values.

\break
\centerline{\bf Acknowledgement} 
\bigskip 

T.$\,$K. would like to thank Koichi Hamaguchi, Taichiro Kugo, Yuuki
Shinbara, and Taizan Watari for very helpful discussions.  He is also
grateful to Jonathan Bagger for kind correspondence. M.Y. would like to
thank Fuminobu Takahashi and Jun'ichi Yokoyama for useful discussions.
The work of T.$\,$K. was supported in part by a Grant-in-Aid
No.~19540268 from the MEXT of Japan.  M.Y. is supported in part by JSPS
Grant-in-Aid for Scientific Research No.~18740157 and No.~19340054.

\vskip .4in


\begin{thebibliography}{99}

\bibitem{Inflation}
See, for example, 
A.~D.~Linde, 
``{\it Particle Physics and Inflationary Cosmology}, ''
(Harwood Academic, 1990) 
[arXiv:hep-th/0503203].\\
A.~R.~Liddle and D.~H.~Lyth, 
``{\it Cosmological Inflation and Large Scale Structure},'' 
(Cambridge University Press, 2000),\\
D.~H.~Lyth and A.~Riotto,
``Particle Physics Models of Inflation and the Cosmological Density
Perturbation,''
Phys.\ Rept.\  {\bf 314}, 1 (1999) 
[hep-ph/9807278].

\bibitem{Komatsu:2008hk}
  E.~Komatsu {\it et al.}  [WMAP Collaboration],
  ``Five-Year Wilkinson Microwave Anisotropy Probe (WMAP)
  Observations:Cosmological Interpretation,''
  [arXiv:0803.0547 [astro-ph]].

\bibitem{Linde:1983gd}
  A.~D.~Linde,
  ``Chaotic Inflation,''
  Phys.\ Lett.\  B {\bf 129}, 177 (1983).

\bibitem{Kawasaki:2000yn}
  M.~Kawasaki, M.~Yamaguchi and T.~Yanagida,
  ``Natural Chaotic Inflation in Supergravity,''
  Phys.\ Rev.\ Lett.\  {\bf 85}, 3572 (2000)
  [arXiv:hep-ph/0004243].

\bibitem{Kawasaki:2000ws}
  M.~Kawasaki, M.~Yamaguchi and T.~Yanagida,
  ``Natural Chaotic Inflation in Supergravity and Leptogenesis,''
  Phys.\ Rev.\  D {\bf 63}, 103514 (2001)
  [arXiv:hep-ph/0011104].

\bibitem{Yamaguchi:2000vm}
  M.~Yamaguchi and J.~Yokoyama,
  ``New Inflation in Supergravity with a Chaotic Initial Condition,''
  Phys.\ Rev.\  D {\bf 63}, 043506 (2001)
  [arXiv:hep-ph/0007021].

\bibitem{Yamaguchi:2001pw}
  M.~Yamaguchi,
  ``Natural Double Inflation in Supergravity,''
  Phys.\ Rev.\  D {\bf 64}, 063502 (2001)
  [arXiv:hep-ph/0103045].

\bibitem{Yamaguchi:2001zh}
  M.~Yamaguchi,
  ``Density Fluctuations and Primordial Black Holes Formation in Natural
  Double Inflation in Supergravity,''
  Phys.\ Rev.\  D {\bf 64}, 063503 (2001)
  [arXiv:hep-ph/0105001].

\bibitem{Goncharov:1983mw}
  A.~B.~Goncharov and A.~D.~Linde,
  ``Chaotic Inflation in Supergravity,''
  Phys.\ Lett.\  B {\bf 139}, 27 (1984).

\bibitem{Goncharov:1985ka}
  A.~S.~Goncharov and A.~D.~Linde,
  ``A Simple Realization of the Inflationary Universe Scenario in SU(1,1)
  Supergravity,''
  Class.\ Quant.\ Grav.\  {\bf 1}, L75 (1984).

\bibitem{Murayama:1993xu}
  H.~Murayama, H.~Suzuki, T.~Yanagida and J.~Yokoyama,
  ``Chaotic Inflation and Baryogenesis in Supergravity,''
  Phys.\ Rev.\  D {\bf 50}, R2356 (1994)
  [arXiv:hep-ph/9311326].

\bibitem{Silverstein:2008sg}
  E.~Silverstein and A.~Westphal,
  ``Monodromy in the CMB: Gravity Waves and String Inflation,''
  Phys.\ Rev.\  D {\bf 78}, 106003 (2008)
  [arXiv:0803.3085 [hep-th]].

\bibitem{KY}
  K.~Kadota and M.~Yamaguchi,
  ``D-term Chaotic Inflation in Supergravity,''
  Phys.\ Rev.\  D {\bf 76}, 103522 (2007), 
  [arXiv:0706.2676 [hep-ph]].

\bibitem{KKY}
  K.~Kadota, T.~Kawano and M.~Yamaguchi,
  ``New D-term chaotic inflation in supergravity and leptogenesis,''
  Phys.\ Rev.\  D {\bf 77}, 123516 (2008)
  [arXiv:0802.0525 [hep-ph]].

\bibitem{Moroi:2005kz}
  T.~Moroi, T.~Takahashi and Y.~Toyoda,
  ``Relaxing Constraints on Inflation Models with Curvaton,''
  Phys.\ Rev.\  D {\bf 72}, 023502 (2005)
  [arXiv:hep-ph/0501007].

\bibitem{Moroi:2005np}
  T.~Moroi and T.~Takahashi,
  ``Implications of the Curvaton on Inflationary Cosmology,''
  Phys.\ Rev.\  D {\bf 72}, 023505 (2005)
  [arXiv:astro-ph/0505339].

\bibitem{Ichikawa:2008iq}
  K.~Ichikawa, T.~Suyama, T.~Takahashi and M.~Yamaguchi,
  Phys.\ Rev.\  D {\bf 78}, 023513 (2008)
  [arXiv:0802.4138 [astro-ph]];
  K.~Ichikawa, T.~Suyama, T.~Takahashi and M.~Yamaguchi,
  Phys.\ Rev.\  D {\bf 78}, 063545 (2008)
  arXiv:0807.3988 [astro-ph].

\bibitem{FI}
  T.~Kawano,
  ``Chaotic D-Term Inflation,''
  arXiv:0712.2351 [hep-th].

\bibitem{WB}
J.~Wess and J.~Bagger,
``{\it Supersymmetry and Supergravity},'' 
(Princeton University, Princeton. NJ, 1992), Second Edition.

\bibitem{Fukugita:1986hr}
  M.~Fukugita and T.~Yanagida,
  ``Baryogenesis without Grand Unification,''
  Phys.\ Lett.\  B {\bf 174}, 45 (1986).

\bibitem{Kumekawa:1994gx}
  K.~Kumekawa, T.~Moroi and T.~Yanagida,
  ``Flat Potential for Inflaton with a Discrete R Invariance in Supergravity,''
  Prog.\ Theor.\ Phys.\  {\bf 92}, 437 (1994)
  [arXiv:hep-ph/9405337].

\bibitem{Lazarides:1999dm}
  G.~Lazarides,
  ``Leptogenesis in Supersymmetric Hybrid Inflation,''
  Springer Tracts Mod.\ Phys.\  {\bf 163}, 227 (2000)
  [arXiv:hep-ph/9904428].

\bibitem{Giudice:1999fb}
  G.~F.~Giudice, M.~Peloso, A.~Riotto and I.~Tkachev,
  ``Production of Massive Fermions at Preheating and Leptogenesis,''
  JHEP {\bf 9908}, 014 (1999)
  [arXiv:hep-ph/9905242].

\bibitem{Asaka:1999yd}
  T.~Asaka, K.~Hamaguchi, M.~Kawasaki and T.~Yanagida,
  ``Leptogenesis in Inflaton Decay,''
  Phys.\ Lett.\  B {\bf 464}, 12 (1999)
  [arXiv:hep-ph/9906366].

\bibitem{Asaka:1999jb}
  T.~Asaka, K.~Hamaguchi, M.~Kawasaki and T.~Yanagida,
  ``Leptogenesis in Inflationary Universe,''
  Phys.\ Rev.\  D {\bf 61}, 083512 (2000)
  [arXiv:hep-ph/9907559].

\bibitem{Kawasaki:2008qe}
  M.~Kawasaki, K.~Kohri, T.~Moroi and A.~Yotsuyanagi,
  ``Big-Bang Nucleosynthesis and Gravitino,''
  Phys.\ Rev.\  D {\bf 78}, 065011 (2008)
  [arXiv:0804.3745 [hep-ph]].

\bibitem{Covi:1996wh}
  L.~Covi, E.~Roulet and F.~Vissani,
  ``CP Violating Decays in Leptogenesis Scenarios,''
  Phys.\ Lett.\  B {\bf 384}, 169 (1996)
  [arXiv:hep-ph/9605319].

\bibitem{Flanz:1994yx}
  M.~Flanz, E.~A.~Paschos and U.~Sarkar,
  ``Baryogenesis from a Lepton Asymmetric Universe,''
  Phys.\ Lett.\  B {\bf 345}, 248 (1995)
  [Erratum-ibid.\  B {\bf 382}, 447 (1996)]
  [arXiv:hep-ph/9411366].

\bibitem{Buchmuller:1997yu}
  W.~Buchmuller and M.~Plumacher,
  ``CP Asymmetry in Majorana Neutrino Decays,''
  Phys.\ Lett.\  B {\bf 431}, 354 (1998)
  [arXiv:hep-ph/9710460].

\bibitem{seesaw}
T. Yanagida, in Proceedings of the Workshop on the Unified Theory and
the Baryon Number of the Universe, Tsukuba, Japan, 1979, edited by
O. Sawada and S. Sugamoto, (KEK, Tsukuba, 1979).

\bibitem{GPS}
M. Gell-Mann, P. Ramond, and R. Slansky,
in ``{\it Supergravity},''
edited by D.Z. Freedman and P. van Nieuwenhuizen 
(North-Holland, Amsterdam, 1979).

\bibitem{Khlebnikov:1988sr}
  S.~Y.~Khlebnikov and M.~E.~Shaposhnikov,
  ``The Statistical Theory of Anomalous Fermion Number Nonconservation,''
  Nucl.\ Phys.\  B {\bf 308}, 885 (1988).

\bibitem{Harvey:1990qw}
  J.~A.~Harvey and M.~S.~Turner,
  ``Cosmological Baryon and Lepton Number in the Presence of Electroweak
  Fermion Number Violation,''
  Phys.\ Rev.\  D {\bf 42}, 3344 (1990).



\end{thebibliography}
\end{document}